\documentclass[twocolumn]{jpsj3}%
\usepackage{bm}
\usepackage{booktabs}
\usepackage{tabularx}
\usepackage{ascmac}
\usepackage{wrapfig}
\usepackage{graphicx}
\usepackage{float}
\usepackage{arydshln}
\usepackage[FIGBOTCAP]{subfigure}
\usepackage{color}

\bmdefine{\bolds}{s}
\bmdefine{\boldi}{i}
\bmdefine{\boldj}{j}
\bmdefine{\boldtau}{\tau}
\bmdefine{\boldsigma}{\sigma}
\bmdefine{\boldk}{k}
\bmdefine{\boldK}{K}
\bmdefine{\boldr}{r}
\bmdefine{\boldj}{j}

\title{Orbital-Selective Superconductivity 
and the Effect of Lattice Distortion\\ 
in Iron-Based Superconductors}

\author{Naoya ARAKAWA$^{1}$\thanks{E-mail address: arakawa@hosi.phys.s.u-tokyo.ac.jp} 
and Masao OGATA$^{1}$$^{,2}$ 
}
\inst{$^{1}$Department of Physics, 
The University of Tokyo,
Tokyo 113-0033, Japan \\
$^{2}$Japan Science and Technology Agency, 
Transformative Research-project on Iron Pnictides,\\ 
5, Sanbancho, Chiyoda, Tokyo 102-0075, Japan
}

\abst{
The superconducting (SC) state of iron-based compounds 
in both tetragonal and orthorhombic phases 
is studied on the basis of 
an effective Hamiltonian composed of 
the kinetic energy including the five Fe $3d$-orbitals, 
the orthorhombic crystalline electric field (CEF) energy, and 
the two-orbital Kugel'-Khomski\u{i}-type 
superexchange interaction.  
Our basic assumption is that 
the antiferromagnetic (AF) state 
in the parent compounds 
can be described by  
the $d_{xz}$ and $d_{yz}$ orbitals, 
and that 
the electrons in these orbitals 
have relatively strong electron correlation 
in the vicinity of the AF state. 
In order to study the physical origin 
of the structure-sensitive 
SC transition temperature, 
the effect of orthorhombic distortion is taken into account 
as the energy-splitting, 
$\Delta_{\textrm{ortho.}}$, 
between the $d_{xz}$ and $d_{yz}$ orbitals. 
We find that 
the eigenvalue of the linearized gap equation  
decreases accompanied with the reduction of 
the partial density of states 
for the $d_{xz}$ and $d_{yz}$ orbitals 
as $\Delta_{\textrm{ortho.}}$ increases, 
and 
that the dominant pairing symmetry is 
an unconventional fully gapped $s_{+-}$-wave pairing. 
We also find large anisotropy of the SC gap function 
in the orthorhombic phase. 
We propose that 
the CEF energy plays an important role in 
controlling $T_{\textrm{c}}$ and the SC gap function, and that 
orbital-selective superconductivity 
is a key feature in iron-based superconductors, 
which causes the structure-sensitive $T_{\textrm{c}}$. 
}

\kword{multiorbital system, 
structure-sensitive superconducting transition temperature, 
iron-based superconductor,  
orbital-selective superconductivity, 
orthorhombic distortion}

\begin{document}
\maketitle

\section{Introduction}
Since the discovery of 
iron-based superconductors,~\cite{Hosono} 
many works have been carried out on the physical properties 
of these new superconductors 
both theoretically and experimentally.~\cite{Fe-review} 
Except for in recent 
studies~\cite{ortho-SC2,11coex}, 
superconductivity has been observed in the tetragonal phase, 
which is the crystalline structure 
above the structural transition temperature 
$T_{\textrm{S}}$.~\cite{Fe-review} 
The crystalline structure becomes orthorhombic 
below $T_{\textrm{S}}$. 
Extensive research 
on these iron-based compounds has clarified 
several general features: 
a structure-sensitive 
superconducting (SC) transition temperature 
$T_{\textrm{c}}$,~\cite{Lee,Mukuda-Tc} 
the existence of disconnected 
Fermi surfaces (FSs),~\cite{ARPES122, deHaas1111, 
deHaas122,deHaas122-P,11-ARPES} 
a small difference between $T_{\textrm{S}}$ and 
the N$\acute{\textrm{e}}$el temperature 
$T_{\textrm{N}}$,~\cite{mag-mom4, 1111TS-TN, 11TS-TN} 
a small magnetic moment 
per Fe atom,~\cite{mag-mom1, mag-mom2, mag-mom4} 
a stripe-type antiferromagnetic (AF) spin 
structure,~\cite{mag-mom1, 122Kitagawa, mag-mom4, NMR-config}  
spin-singlet pairing,~\cite{122Kitagawa, NMR-singlet0,
NMR-singlet1,NMR-singlet2,
NMR-singlet3} 
various power-law dependences on the temperature 
of the spin-lattice relaxation rate 
$T_{1}^{-1}$ below $T_{\textrm{c}}$,~\cite{122Kitagawa, NMR-singlet0,
NMR-singlet1,NMR-singlet2, 
NMR-singlet3, mag-mom3, NMR1, NMR2, K122NMR, 
Co122NMR, P122NMR, 11NMR}  
the absence of the Hebel-Slichter (HS) peak 
just below $T_{\textrm{c}}$,~\cite{122Kitagawa, NMR-singlet0,
NMR-singlet1,NMR-singlet2, 
NMR-singlet3, mag-mom3, NMR1, NMR2, K122NMR, 
Co122NMR, P122NMR, 11NMR} 
and the robustness of 
$T_{\textrm{c}}$ against nonmagnetic 
impurities.~\cite{Co-imp-Sato, Ru-imp-Sato, 122-imp} 
Note that 
the resistivity above $T_{\textrm{N}}$ is metallic;~\cite{Fe-review} 
thus, it is appropriate to regard the magnetically ordered state 
a spin-density-wave state. 
For simplicity, 
we call this magnetically ordered state an AF state 
in the following. 
The phase diagrams of 1111, 122, 11, 
and 111 systems are similar. 
The only difference is the following:  
in the 1111 system, 
the SC phase and AF phase are separated,~\cite{Hosono} 
while the two phases adjoin each other 
in the other systems.~\cite{122PD,correlation-MS2,11PD} 

In early theoretical studies,~\cite{Mazin, Kuroki} 
it was proposed that 
the symmetry of the SC gap function is 
an unconventional $s_{+-}$-wave. 
Several experimental results have supported this proposition, 
for example, 
the existence of a resonance peak 
in neutron scattering,~\cite{neutron-res} 
the absence of the HS peak of $T_{1}^{-1}$ 
just below $T_{\textrm{c}}$,~\cite{122Kitagawa, NMR-singlet0,
NMR-singlet1,NMR-singlet2, 
NMR-singlet3, mag-mom3, NMR1, NMR2, K122NMR, 
Co122NMR, P122NMR, 11NMR} 
its revival in an overdoped compound,~\cite{Mukuda-coherence} 
and 
quasi-particle interference patterns.~\cite{11QPI} 
On the other hand, 
experiments on the nonmagnetic impurity 
effect~\cite{Co-imp-Sato, Ru-imp-Sato, 122-imp} 
indicate that the iron-based superconductors are robust 
against nonmagnetic impurities. 
This is inconsistent with the above theoretical predictions, 
since some theoretical studies on 
the impurity effect based on a standard \textit{T}-matrix 
approximation 
using realistic parameters~\cite{Kontani-imp} have shown 
that $T_{\textrm{c}}$ with the $s_{+-}$-wave symmetry 
decreases more rapidly than 
the experimental results~\cite{Co-imp-Sato, 
Ru-imp-Sato,122-imp}. 
Although it has been proposed~\cite{Kontani1, Yanagi} that 
the orbital fluctuation plays an important role 
in explaining this discrepancy, 
the pairing mechanism of iron-based superconductors 
is still controversial. 

We consider that the key to clarifying 
the pairing mechanism 
is to understand the role of lattice distortion 
in determining the electronic states 
since the unusual properties will be 
structure-sensitive. 
In this sense, 
the $d_{xz}$ and $d_{yz}$ orbitals will be more important 
than the other Fe $3d$ orbitals 
since they are closely connected to the orthorhombic distortion. 
The results of several recent experiments 
for the underdoped 122 system 
have suggested the importance 
of the orbital degree of freedom and lattice distortion: 
anisotropic in-plane resistivity above $T_{\textrm{S}}$~\cite{aniso-resis}, 
anisotropic band dispersion 
above $T_{\textrm{S}}$~\cite{Shen}, 
the large incoherent electronic Raman spectrum 
for the $d_{xz}$ and $d_{yz}$ orbitals~\cite{122Raman-loc}, 
the smaller mobility of holes observed in 
the Hall resistivity measurement using a two-carrier model~\cite{122Hall-loc}, 
the orbital-dependent modification of 
the electronic states 
across $T_{\textrm{N}}$ observed 
in the angle-resolved photoemission spectroscopy (ARPES) 
measurement~\cite{Shimojima}, 
the marked softening of the elastic constants 
at $T_{\textrm{S}}$ observed in 
ultrasound measurements~\cite{ultrasonic-122-0,ultrasonic-122-1}, 
the good correlation between 
the orthorhombic distortion and 
the AF order observed in neutron and X-ray 
measurements~\cite{correlation-MS,
correlation-MS2}. 
It is probable that 
the AF state in the parent compounds 
can be described by 
the $d_{xz}$ and $d_{yz}$ orbitals.~\cite{FO-aniso1, 
KuboAF, KuboAF2} 
In both parent and underdoped 122 and 11 systems, 
a large enhancement of the effective mass 
has been observed 
in several experiments.~\cite{
deHaas122, deHaas122-P, 
ARPES122,11-ARPES,Te-deHaas122,122PES, 122deHaas2, 
11PES, 11PES2, optical-122} 

From the above experimental results, 
we expect that 
the electron correlation 
in the $d_{xz}$ and $d_{yz}$ orbitals 
will play an important role 
in both parent and underdoped compounds. 
Although there have been several 
theoretical proposals~\cite{FO-aniso1,FO-aniso2,FO-aniso3,FO-aniso4} 
that 
a coupled spin and orbital order occurs 
in the AF state, 
the effect of this coupling on the SC state 
has not been clarified yet. 

In this paper, 
we focus on the SC state 
in both tetragonal 
and orthorhombic phases 
in order to investigate 
the effect of orthorhombic distortion 
on $T_{\textrm{c}}$ 
and 
the role of the spin-orbital coupling 
in determining the SC state. 
We assume that 
relatively strong electron correlation 
exists for the $d_{xz}$ and $d_{yz}$ orbitals, 
and that the superconductivity is induced 
from the Kugel'-Khomski\u{i} (KK)-type 
superexchange interaction~\cite{K-K} 
for the $d_{xz}$ and $d_{yz}$ orbitals. 
We call this orbital-selective superconductivity 
in the sense that 
only the $d_{xz}$ and $d_{yz}$ orbitals induce superconductivity.  
For this purpose, 
we introduce an effective model 
composed of the kinetic energy 
including the five Fe $3d$-orbitals, 
the orthorhombic crystalline electric field (CEF) energy, 
and 
the KK-type superexchange interaction. 
There have been some previous studies 
that discuss the SC state 
induced from the superexchange interaction.~\cite{tJ} 
In the present paper, 
a more realistic model Hamiltonian is considered. 
We use the two-dimensional (2-D) tight-binding model 
of the five Fe $3d$ orbitals 
downfolded from the local density approximation (LDA) 
calculation.~\cite{Kuroki} 
In order to study the effect of orthorhombic distortion, 
we introduce the splitting of energy levels, 
$\Delta_{\textrm{ortho.}}$, 
between the $d_{xz}$ and $d_{yz}$ orbitals. 
Furthermore, 
we will include the effect of Coulomb interaction 
by a band renormalization 
in which the total bandwidth 
of the five Fe $3d$-orbitals and the hybridizations 
concerning the $d_{xz}$ and $d_{yz}$ orbitals 
is reduced. 
This procedure is based on the same principle as 
the Gutzwiller approximation.~\cite{Gutzwiller} 
From the experimental result~\cite{deHaas122, deHaas122-P, 
ARPES122,11-ARPES,Te-deHaas122,deHaas1111,122PES, 122deHaas2, 
11PES, 11PES2, optical-122,optical-1111} 
that the effective mass is at least twice the band mass, 
we use a renormalized band whose width is nearly 
half that of the original one. 

We study the behavior of the eigenvalue, 
$\lambda_{\textrm{e}}$, of 
the linearized gap equation 
and the pairing symmetry 
for $0 \leq \Delta_{\textrm{ortho.}} \leq 0.1$ eV. 
We find that 
$\lambda_{\textrm{e}}$ decreases 
as $\Delta_{\textrm{ortho.}}$ increases, 
and that 
the SC state 
in the tetragonal phase 
gives the maximum $T_{\textrm{c}}$. 
The former behavior is accompanied 
by the reduction of the partial density of states (pDOS) 
for the $d_{xz}$ and $d_{yz}$ orbitals. 
We also find that 
the dominant pairing symmetry 
is a fully gapped $s_{+-}$-wave pairing 
both in tetragonal and orthorhombic phases, 
and 
the second most dominant pairing symmetry 
is a $d_{xy}$-wave pairing whose 
$\lambda_{\textrm{e}}$ rapidly decreases 
as $\Delta_{\textrm{ortho.}}$ increases. 
The finite hybridizations between 
the strongly correlated orbitals and 
the weakly correlated orbitals 
cause the change in the pDOS 
for the $d_{xz}$ and $d_{yz}$ orbitals 
near the Fermi level. 
This is one of the physical origins 
of the sensitivity of $T_{\textrm{c}}$. 
Large anisotropy of the SC gap function is found 
in the orthorhombic phase. 

The outline of this paper is as follows: 
In \S 2, 
we explain the form of the kinetic energy 
and 
the method used to include 
the effect of orthorhombic distortion, 
and we derive the KK-type superexchange interaction  
for the $d_{xz}$ and $d_{yz}$ orbitals. 
In order to discuss the SC state 
for the spin-singlet pairing, 
we use the mean field approximation (MFA) 
in our effective model 
and 
derive the linearized gap equation. 
\S 3 is devoted to showing the results 
of the eigenvalue 
and the pairing symmetry 
for various values of $\Delta_{\textrm{ortho.}}$. 
In \S 4, 
we address the physical meaning of the obtained results, 
compare them with other previous theories, and 
discuss their correspondence with previous experimental results. 
The paper concludes with a summary of our results in \S 5. 

\section{Formalism}
In this section, 
we explain the methods used to take account of  
the effect of orthorhombic distortion 
and to construct an effective Hamiltonian, 
$H_{\textrm{eff}}$, for the discussion of 
iron-based superconductors 
in both tetragonal and orthorhombic phases, 
assuming that 
the $d_{xz}$ and $d_{yz}$ orbitals have relatively 
strong electron correlation. 
The effective Hamiltonian is  
\begin{align}
H_{\textrm{eff}}=\tilde{H}_{0}+H_{\textrm{ortho.}}
+H_{\textrm{int}} \ ,
\end{align}
where $\tilde{H}_{0}$ is the kinetic energy 
of the tetragonal phase 
modified by band renormalization, 
the detail of which is described in \S 2.1, 
$H_{\textrm{ortho.}}$ is the orthorhombic CEF energy, 
and $H_{\textrm{int}}$ is the effective interaction. 
In the following, 
the coordinates $x$ and $y$ 
are chosen in 
the directions of the 
nearest-neighbor (n.n.) Fe atoms 
in a unit cell. 
Note that, 
in the orthorhombic phase, 
the $a$-axis corresponds to 
the $x$-direction. 
For convenience, 
the five Fe $3d$-orbitals, 
$d_{xz}$, $d_{yz}$, $d_{xy}$, 
$d_{x^{2}-y^{2}}$, and $d_{z^{2}}$, 
are labeled 
1, 2, 3, 4, and 5, 
respectively. 
We define the band filling $n_{\textrm{e}}$ 
as the number of electrons per site. 
In iron-based compounds, 
$n_{\textrm{e}}$ is $6+x$, 
where $x$ represents the doping level.

\subsection{Kinetic energy and orthorhombic CEF energy}
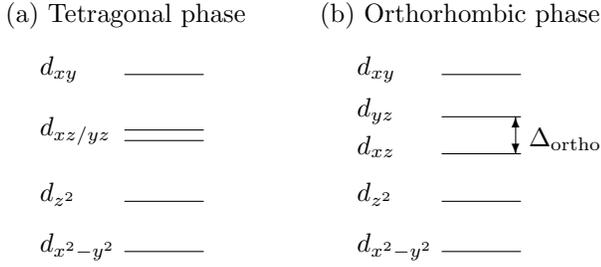
\begin{figure}[tb]
\begin{picture}(100,100)
\put(20,3){$d_{x^{2}-y^{2}}$}
\put(52,3){\line(1,0){30}}
\put(20,22){$d_{z^{2}}$}
\put(52,22){\line(1,0){30}}
\put(20,47){$d_{xz/yz}$}
\put(52,45){\line(1,0){30}}
\put(52,49){\line(1,0){30}}
\put(20,70){$d_{xy}$}
\put(52,70){\line(1,0){30}}
\put(7,90){(a) Tetragonal phase}
\put(140,3){$d_{x^{2}-y^{2}}$}
\put(172,3){\line(1,0){30}}
\put(140,22){$d_{z^{2}}$}
\put(172,22){\line(1,0){30}}
\put(140,40){$d_{xz}$}
\put(172,40){\line(1,0){30}}
\put(200,47){\vector(0,1){7}}
\put(200,47){\vector(0,-1){7}}
\put(140,54){$d_{yz}$}
\put(172,54){\line(1,0){30}}
\put(205,43){$\Delta_{\textrm{ortho.}}$}
\put(140,70){$d_{xy}$}
\put(172,70){\line(1,0){30}}
\put(126,90){(b) Orthorhombic phase}
\end{picture}
\vspace{4pt}
\caption{
Schematic energy levels 
for the five Fe $3d$ orbitals 
in the (a) tetragonal and 
(b) orthorhombic phases. }
\label{fig:levels-both}
\vspace{-5pt}
\end{figure} 
In order to describe the electronic states 
of iron-based superconductors, 
we use the 2-D five-orbital model 
downfolded from 
the LDA calculation,~\cite{Kuroki} 
\begin{align}
H_{0}
=& \ 
\textstyle\sum\limits
_{\boldi, \boldj}
\textstyle\sum\limits_{a,b=1}^{5}
\textstyle\sum\limits_{\sigma}
( t_{ab}^{\boldi, \boldj} 
c^{\dagger}_{\boldi a \sigma} 
c_{\boldj b \sigma} 
+ \textrm{h.c.} )\notag\\
+& 
\textstyle\sum\limits
_{\boldi}
\textstyle\sum\limits_{a=1}^{5}
\textstyle\sum\limits_{\sigma} 
( E_{a} 
- \mu ) 
n_{\boldi a \sigma} \ ,\notag\\ 
=& \ 
\textstyle\sum\limits
_{\boldk}
\textstyle\sum\limits_{a,b=1}^{5}
\textstyle\sum\limits_{\sigma} 
\epsilon_{ab}(\boldk ) 
c^{\dagger}_{\boldk a \sigma} 
c_{\boldk b \sigma} \ ,  \label{eq:bare-dis}
\end{align}
where $c^{\dagger}_{\boldi a \sigma}$ 
($c_{\boldi a \sigma}$) 
is the creation (annihilation) operator that 
creates (annihilates) an electron in orbital $a$ 
with spin $\sigma$ at site $\boldi$, 
h.c. means the Hermitian conjugate, 
$n_{\boldi a \sigma}=
c^{\dagger}_{\boldi a \sigma} 
c_{\boldi a \sigma}$, 
$c_{\boldk a \sigma}$ is the Fourier component of 
$c_{\boldi a \sigma}$, 
and 
$t_{ab}^{\boldi, \boldj}$, $E_{a}$, 
and $\epsilon_{ab}(\boldk)$ 
denote the values of in-plane hopping integrals, 
on-site energies,  
and 
energy dispersions measured from 
the chemical potential $\mu$, 
respectively. 

The effect of electron correlation 
is taken into account 
by modifying $H_{0}$ to 
\begin{align}
\tilde{H}_{0}
=& \ 
\textstyle\sum\limits
_{\boldk}
\textstyle\sum\limits_{a,b=1}^{5}
\textstyle\sum\limits_{\sigma} 
\tilde{\epsilon}_{ab}(\boldk ) 
c^{\dagger}_{\boldk a \sigma} 
c_{\boldk b \sigma} \ , \label{eq:mod-hop}
\end{align}
where $\tilde{\epsilon}_{ab}(\boldk)$ 
is the renormalized energy dispersion. 
$\tilde{\epsilon}_{ab}(\boldk )$ is constructed as follows. 
All $t_{ab}^{\boldi, \boldj}$ 
in eq. (\ref{eq:bare-dis}) are reduced 
by a factor of $0.6$, 
assuming band renormalization 
due to the electron correlation effect. 
In addition to this, 
$t_{1a}^{\boldi, \boldj}$ 
and $t_{2a}^{\boldi, \boldj}$ ($a=$1 $-$ 5) 
are further reduced by a factor of $0.75$. 
We use this modification assuming a relatively 
strong electron correlation in the $d_{xz}$ and $d_{yz}$ orbitals.  
Finally, the on-site energies, $E_{a}$ ($a=$1 $-$ 5), 
are adjusted as 
\begin{align}
E_{a} \rightarrow
\begin{cases}
E_{a} 
+ 0.0045 \ (\textrm{eV})  \ \ \ 
\textrm{for} \ \ a=1,2,3 \  
\\  
E_{a} - 0.03 \ (\textrm{eV}) \ \ \ \ \ \ 
\textrm{for} \ \ a=4,5
\end{cases} \label{eq:Ea-mod} 
\end{align}
in order to make the total occupation number 
of the $d_{xz}$ and $d_{yz}$ orbitals, 
$(n_{\textrm{e}})_{xz+yz}$, close to $3$, 
since we assume a KK-type interaction 
as discussed below. 
Note that the center of the energy levels remains 
zero under the transformation in eq. (\ref{eq:Ea-mod}). 
 
In order to include the effect 
of orthorhombic distortion, 
we use 
the orthorhombic CEF energy 
for the $d_{xz}$ and $d_{yz}$ 
orbitals (i.e., $a=1$ and $2$), 
\begin{align}
H_{\textrm{ortho.}}=
\textstyle\sum\limits_{\boldk} 
\textstyle\sum\limits_{\sigma} 
\dfrac{\Delta_{\textrm{ortho.}}}{2} 
( n_{\boldk 2 \sigma} 
- n_{\boldk 1 \sigma} ), \label{eq:ortho.}
\end{align}
where $\Delta_{\textrm{ortho.}}$ 
is the splitting 
between the energy levels of 
the $d_{xz}$ and $d_{yz}$ 
orbitals. 
It has been experimentally shown that 
the $a$-axis is longer than 
the $b$-axis 
in the orthorhombic phase.~\cite{Fe-review} 
Since As ions have negative charge 
and the $d_{xz}$ orbital extends 
toward As ions, 
we expect that 
the energy level of the $d_{xz}$ orbital 
is lower than that of the $d_{yz}$ orbital, 
i.e., 
\begin{align}
E_{1} \rightarrow  E_{1} 
- \frac{ \Delta_{\textrm{ortho.}}}{2} , \
E_{2} \rightarrow  E_{2} 
+ \frac{ \Delta_{\textrm{ortho.}}}{2} .
\end{align}
Note that this expectation is consistent with 
the result of 
an ARPES measurement.~\cite{Shen} 
Schematic energy levels for the five Fe $3d$ orbitals 
are shown in Fig. \ref{fig:levels-both}. 

In this work, 
we use the value of $\Delta_{\textrm{ortho.}}$ 
as a parameter ranging from $0$ to $0.1$ eV. 
According to the ARPES measurement,~\cite{Shen} 
$\Delta_{\textrm{ortho.}}$ is approximately equal 
to $0.06$ eV at $T/T_{\textrm{N}}\sim 0.58$ 
for $\textrm{Ba}\textrm{Fe}_{2}\textrm{As}_{2}$. 
Moreover, 
the LDA calculation 
for the parent compounds~\cite{Pickett} 
indicates that 
$\Delta_{\textrm{ortho.}}$ is on the order of $0.1$ eV. 
In order to maintain the topology of the FS, 
we choose $\Delta_{\textrm{ortho.}}$ so as to satisfy the inequalities 
$\Delta_{\textrm{ortho.}}\leq 2(E_{1} - E_{5})$ 
and 
$\Delta_{\textrm{ortho.}}\leq 2(E_{3} - E_{2})$. 
Precisely speaking, 
the orthorhombic distortion also leads to 
anisotropies in the hopping integrals. 
However, 
we neglect this effect in the following calculations, 
the validity of which will be discussed in \S 4. 

$\tilde{H}_{0}+H_{\textrm{ortho.}}$ is diagonalized as 
\begin{align}
\tilde{H}_{0}+H_{\textrm{ortho.}}
=& \ 
\textstyle\sum\limits
_{\boldk}
\textstyle\sum\limits_{\alpha}
\textstyle\sum\limits_{\sigma} 
\tilde{\epsilon}_{\alpha}(\boldk ) 
c^{\dagger}_{\boldk \alpha \sigma} 
c_{\boldk \alpha \sigma} \ ,  
\end{align}
where $\alpha$ is the band index and 
$\tilde{\epsilon}_{\alpha}(\boldk )$ 
is the renormalized energy dispersion for band $\alpha$. 
Here, $c_{\boldk \alpha \sigma}$ is related 
to $c_{\boldk a \sigma}$ by the unitary transformation 
\begin{align}
c_{\boldk a \sigma}=
\textstyle\sum\limits_{\alpha}
(U_{\boldk})_{a \alpha} c_{\boldk \alpha \sigma} \ 
.\label{eq:unitary}
\end{align}

\subsection{KK-type superexchange interaction}
\begin{figure}[tb]
\vspace{8pt}
\hspace{15pt}
\begin{picture}(60,60)
\put(-18,57){(a) $U$}
\put(-14,17){\line(1,0){12}}
\put(-10,11){\vector(0,1){12}}
\put(-5,23){\vector(0,-1){12}}
\put(-14,40){\line(1,0){12}}
\put(1,26){or}
\put(14,17){\line(1,0){12}}
\put(18,34){\vector(0,1){12}}
\put(22,46){\vector(0,-1){12}}
\put(14,40){\line(1,0){12}}
\put(32,6){\dashbox{0.8}(0,60)}
\linethickness{0.16pt}
\qbezier(56,11)(50,28)(56,45)
\thinlines
\put(35,57){(b) $U^{\prime}$}
\put(38,25){$\frac{1}{\sqrt{2}}$}
\put(57,17){\line(1,0){12}}
\put(62,23){\vector(0,-1){12}}
\put(57,40){\line(1,0){12}}
\put(64,34){\vector(0,1){12}}
\put(71,26){$-$}
\put(80,17){\line(1,0){12}}
\put(85,11){\vector(0,1){12}}
\put(80,40){\line(1,0){12}}
\put(86,46){\vector(0,-1){12}}
\linethickness{0.16pt}
\qbezier(93,11)(99,28)(93,45)
\thinlines
\put(102,6){\dashbox{0.8}(0,60)}
\put(105,57){(c) $U^{\prime}-J_{\textrm{H}}$}
\put(110,17){\line(1,0){12}}
\put(116,11){\vector(0,1){12}}
\put(118,34){\vector(0,1){12}}
\put(110,40){\line(1,0){12}}
\put(125,26){or}
\linethickness{0.16pt}
\qbezier(156,11)(150,28)(156,45)
\thinlines
\put(137,25){$\frac{1}{\sqrt{2}}$}
\put(157,17){\line(1,0){12}}
\put(163,23){\vector(0,-1){12}}
\put(157,40){\line(1,0){12}}
\put(163,34){\vector(0,1){12}}
\put(171,26){$+$}
\put(180,17){\line(1,0){12}}
\put(186,11){\vector(0,1){12}}
\put(180,40){\line(1,0){12}}
\put(188,46){\vector(0,-1){12}}
\linethickness{0.16pt}
\qbezier(194,11)(200,28)(194,45)
\thinlines
\put(202,26){or}
\put(214,17){\line(1,0){12}}
\put(221,23){\vector(0,-1){12}}
\put(221,46){\vector(0,-1){12}}
\put(214,40){\line(1,0){12}}
\end{picture}
\vspace{-14pt}
\caption{
Schematic diagrams of 
possible intermediate states 
on a single site 
and 
the corresponding energies 
neglecting the effect of $\Delta_{\textrm{ortho.}}$. 
The upper and lower horizontal lines 
represent the $d_{yz}$ and $d_{xz}$ orbitals, 
respectively. 
Up (down) arrows correspond to 
spin-up (spin-down) electrons. }
\vspace{-5pt}
\label{fig:inter}
\end{figure}
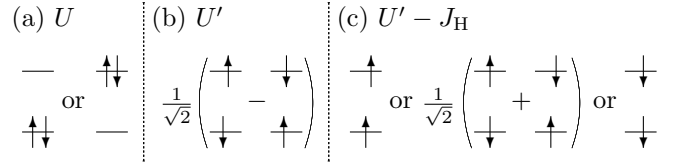
We derive the KK-type superexchange interaction 
between the $d_{xz}$ and $d_{yz}$ orbitals 
in order to discuss the electronic states 
of iron-based superconductors. 
We assume that 
relatively strong electron correlation 
exists in the $d_{xz}$ and $d_{yz}$ orbitals 
since the AF state 
is realized mainly in these orbitals.~\cite{FO-aniso1, 
KuboAF, KuboAF2} 

First, 
we assume standard on-site interactions 
for the $d_{xz}$ and $d_{yz}$ orbitals, 
i.e., 
the intraorbital Coulomb interaction $U$, 
the interorbital Coulomb interaction $U^{\prime}$, 
and the Hund's rule coupling $J_{\textrm{H}}$. (The 
pair-hopping term does not directly affect 
the electronic state for the case with 
$(n_{\textrm{e}})_{xz+yz}=1$ or $3$ 
in the strong coupling limit.)
In this work, 
$U,\ U^{\prime}>J_{\textrm{H}}$ is assumed. 

Assuming that $(n_{\textrm{e}})_{xz+yz}$ is close to 3 or 1 and 
using the second-order perturbation theory 
from the strong coupling limit, 
we obtain 
\begin{align}
&H_{\textrm{int}}
=
\textstyle\sum\limits_{\boldi,\sigma}
\textstyle\sum\limits_{\boldj=\textrm{n.n.}}
\textstyle\sum\limits_
{a, a^{\prime}, b=1}^{2} 
\dfrac{t_{ab} t_{ba^{\prime}}}{U} 
c_{\boldi a^{\prime} \sigma}^{\dagger} 
c_{\boldj b \bar{\sigma}}^{\dagger} 
(
c_{\boldj b \sigma} 
c_{\boldi a \bar{\sigma}} 
- 
c_{\boldj b \bar{\sigma}} 
c_{\boldi a \sigma}
)\notag\\
+& 
\textstyle\sum\limits_{\boldi,\sigma}
\textstyle\sum\limits_{\boldj=\textrm{n.n.}}
\textstyle\sum\limits_
{a, a^{\prime}, b, b^{\prime}=1}^{2}  
\dfrac{t_{ab} t_{b^{\prime}a^{\prime}}}{2 U^{\prime} } 
c_{\boldi a^{\prime} \sigma}^{\dagger} 
c_{\boldj \bar{b^{\prime}} \bar{\sigma}}^{\dagger} 
( 
c_{\boldj \bar{b} \sigma} 
c_{\boldi a \bar{\sigma}} 
- 
c_{\boldj \bar{b} \bar{\sigma}} 
c_{\boldi a \sigma} 
)\notag\\
+& 
\textstyle\sum\limits_{\boldi,\sigma}
\textstyle\sum\limits_{\boldj=\textrm{n.n.}}
\textstyle\sum\limits_
{a, a^{\prime}, b, b^{\prime}=1}^{2} 
(-1)^{1+b+b^{\prime}} 
\dfrac{t_{ab} t_{b^{\prime}a^{\prime}}}
{2 (U^{\prime}-J_{\textrm{H}}) } 
c_{\boldi a^{\prime} \sigma}^{\dagger} 
c_{\boldj \bar{b^{\prime}} \bar{\sigma}}^{\dagger}\notag\\
(& 
c_{\boldj \bar{b} \sigma} 
c_{\boldi a \bar{\sigma}} 
+ 
c_{\boldj \bar{b} \bar{\sigma}} 
c_{\boldi a \sigma} 
)\notag\\
+& 
\textstyle\sum\limits_{\boldi,\sigma}
\textstyle\sum\limits_{\boldj=\textrm{n.n.}}
\textstyle\sum\limits_
{a, a^{\prime}, b, b^{\prime}=1}^{2} 
(-1)^{1+b+b^{\prime}} 
\dfrac{t_{ab} t_{b^{\prime}a^{\prime}}}
{U^{\prime}-J_{\textrm{H}}} 
c_{\boldi a^{\prime} \sigma}^{\dagger} 
c_{\boldj \bar{b^{\prime}} \sigma}^{\dagger}  
c_{\boldj \bar{b} \sigma} 
c_{\boldi a \sigma} \notag\\
+& 
( \textrm{n.n.n. terms} ) 
, \label{eq:KK}
\end{align}
where $a$ and $\sigma$ represent 
orbital and spin degrees of freedom, 
respectively, 
$\boldj=$n.n. denotes 
the summation over 
the n.n. sites of $\boldi$, 
$\bar{a}$ represents the other orbital 
from $a$, and 
$\bar{\sigma}=-\sigma$. 
Possible intermediate states and 
corresponding energies are shown in Fig. \ref{fig:inter}. 
In eq. (\ref{eq:KK}), 
we have also included the superexchange interactions 
originating from the next-nearest-neighbor (n.n.n.) hoppings, 
since they are not negligible 
in iron-based compounds. 
Actually, 
the magnitudes of the hopping integrals 
are strongly affected 
by hybridizations between the $2p$ orbitals of As 
(or P, Te, or Se) 
and the $3d$ orbitals of Fe. 

When $(n_{\textrm{e}})_{xz+yz}$ is away from 3 or 1, 
many other superexchange interaction terms appear. 
In the doped case, 
these terms may affect the properties of the SC state. 
However, we neglect these effects as far as 
$(n_{\textrm{e}})_{xz+yz}$ is close to $3$ or $1$. 
Furthermore, 
we have neglected the effect of 
$\Delta_{\textrm{ortho.}}$ in the denominators 
in eq. (\ref{eq:KK}) since we assume 
$\Delta_{\textrm{ortho.}} << 
U, U^{\prime}, 
(U^{\prime}-J_{\textrm{H}})$, 
with the latter being on the order of $1$ eV. 
Note that we do not use the pseudospin and spin 
operators in the KK-type interactions 
since we intend to study the SC states in the doped cases. 

We consider that the KK-type superexchange interaction 
in eq. (\ref{eq:KK}) can describe the stripe-type AF state 
when there is no doping or when $(n_{\textrm{e}})_{xz+yz}=3$. 
When an orbital-ferromagnetic phase transition occurs, 
our effective interaction becomes similar to 
the anisotropic Heisenberg spin Hamiltonian with 
n.n. and n.n.n. interactions.~\cite{loc-AF1,loc-AF2} 
In this case, 
the stripe-type AF state can be described naturally. 
Furthermore, 
the tight-binding calculation for the 1111, 122, and 111 systems 
shows that 
the magnetic interactions are short-range.~\cite{As-position-mAF} 
This is consistent with our KK-type interaction. 

\subsection{Superconductivity}
As discussed in \S 1, 
we assume that the KK-type superexchange interaction 
for the $d_{xz}$ and $d_{yz}$ orbitals 
in eq. (\ref{eq:KK}) induces superconductivity, 
and thus apply the MFA to these interaction terms. 
Since the terms 
with a denominator of 
$(U^{\prime}-J_{\textrm{H}})$ in eq. (\ref{eq:KK}) 
give no contribution to 
the spin-singlet pairing, 
we consider only 
the terms with $U$ and $U^{\prime}$ 
in the denominator. 
Note that 
there are hybridizations between the five Fe $3d$-orbitals, 
while the superexchange interactions are only between the $d_{xz}$ and 
$d_{yz}$ orbitals. 

After a straightforward calculation, 
the effective Hamiltonian within the MFA 
is obtained as 
\begin{align}
H_{\textrm{eff}}^{\textrm{MFA}}
=& \ 
\textstyle\sum\limits_{\boldk}
\textstyle\sum\limits_{a,b=1}^{5}
\textstyle\sum\limits_{\sigma}
\tilde{\epsilon}_{ab}(\boldk) c^{\dagger}_{\boldk a \sigma} 
c_{\boldk b \sigma}\notag\\
+& 
\textstyle\sum\limits_{\boldk} 
\textstyle\sum\limits_{\sigma} 
\dfrac{\Delta_{\textrm{ortho.}}}{2} 
( n_{\boldk 2 \sigma} 
- n_{\boldk 1 \sigma} )\notag\\
+& 
\textstyle\sum\limits_{\boldk}
\textstyle\sum\limits_{a,b=1}^{2}
( \Delta_{ab}(\boldk) 
c_{\boldk a \uparrow}^{\dagger} 
c_{-\boldk b \downarrow}^{\dagger} 
+ \textrm{h.c.} 
) . \label{eq:Heff}
\end{align}
Here, 
$\Delta_{ab}(\boldk)$ is 
the SC gap function 
for the $d_{xz}$ and $d_{yz}$ orbitals 
defined as
\begin{align}
\Delta_{ab}(\boldk)
=&  
\frac{1}{N} 
\textstyle\sum\limits_{\boldk^{\prime}}  
\textstyle\sum\limits_{a^{\prime}, b^{\prime}=1}^{2}
V_{a a^{\prime}, b b^{\prime}}(\boldk - \boldk^{\prime})
\langle 
c_{-\boldk^{\prime} a^{\prime} \downarrow} 
c_{\boldk^{\prime} b^{\prime} \uparrow} 
\rangle , \label{eq:defDel1}
\end{align}
with $N$ being the number of lattice sites of Fe atoms. 
For simplicity of the numerical calculation, 
we introduce the cutoff energy 
$\varepsilon_{\textrm{c}}$, i.e., the $\boldk$-sum is 
restricted. 
Here, 
$V_{a a^{\prime}, b b^{\prime}}(\boldk - \boldk^{\prime})$ 
is the coefficient of the two-body interaction 
between ($\boldk$ $a$ $\uparrow$, 
$-\boldk$ $b$ $\downarrow$) electrons 
and ($\boldk^{\prime}$ $a^{\prime}$ $\uparrow$, 
$-\boldk^{\prime}$ $b^{\prime}$ $\downarrow$) 
electrons, 
which is obtained from eq. (\ref{eq:KK}) 
by Fourier transformation. 
Explicitly, 
$V_{a^{\prime} a^{\prime\prime}, b^{\prime} b^{\prime\prime}}
(\boldk - \boldk^{\prime})$ 
is written as 
\begin{align}
V_{a^{\prime} a^{\prime\prime}, b^{\prime} b^{\prime\prime}}
(\boldk - \boldk^{\prime})
=
 J_{a^{\prime} a^{\prime\prime}, b^{\prime} b^{\prime\prime}} 
f(\boldk - \boldk^{\prime}), 
\end{align}
where $f(\boldK \equiv \boldk - \boldk^{\prime})$ 
and $J_{a^{\prime} a^{\prime\prime}, b^{\prime} b^{\prime\prime}}$ 
are defined as
\begin{align}
f(\boldK)
=& 
\begin{cases}
2 \cos( K_{x} )
\ \ \ \ 
\text{for} \ 
( \boldi - \boldj ) 
= ( 1, 0) \ \text{and} \ ( -1, 0)\\
2 \cos( K_{y} )
\ \ \ \
\text{for} \ 
( \boldi - \boldj ) 
= (0, 1) \ \text{and} \ (0, -1)\\ 
4 \cos( K_{x} ) \cos( K_{y} )
\ \ \ \
\text{for $( \boldi - \boldj )=$ n.n.n.}  \\
\end{cases},
\end{align} 
and 
\begin{align}
&J_{a^{\prime} a^{\prime\prime}, b^{\prime} b^{\prime\prime}}\notag\\
=& 
\begin{cases}
\ \ \ \ \dfrac{4 t_{aa}^{2}}{U} 
+ \dfrac{2 t_{12}^{2}}{U^{\prime}} 
\ \ \ \  
\ \ \ \ \ \ \ \ 
\text{for} \ 
a^{\prime}=b^{\prime}
=a^{\prime\prime}
=b^{\prime\prime}
=a \\
\dfrac{(t_{11}^{2} + t_{22}^{2})}{ U^{\prime} } 
+\dfrac{4 t_{12}^{2}}{U}
\ \ \ \ \ \ \  \  
\text{for} \ 
a^{\prime}=
a^{\prime\prime}=a, 
b^{\prime}=
b^{\prime\prime}=\bar{a} \\
\ \ \ \ \ \ \ \ \
\dfrac{2 t_{12}^{2}}{U^{\prime}} 
\ \ \ \ \ \ \ \ 
\ \ \ \ \ \ \ \ \ 
\text{for} \ 
a^{\prime}=
b^{\prime}=a, 
a^{\prime\prime}=
b^{\prime\prime}=
\bar{a} \\ 
\ \ \ \ \ \ \ 
\dfrac{2 t_{11} t_{22}}{U^{\prime}} 
\ \ \ \ \ \ \ \ \ \ \ \ \ \  \ 
\text{for} \ 
a^{\prime}=
b^{\prime\prime}=a, 
b^{\prime}=
a^{\prime\prime}=\bar{a} \\
\dfrac{2 t_{aa} t_{12}}{U} 
+ \dfrac{ t_{12} (t_{11}+t_{22}) }{U^{\prime}} 
\ 
\text{for} \ 
a^{\prime\prime}=
b^{\prime\prime}= 
b^{\prime}=a, 
a^{\prime}=\bar{a} \\
\dfrac{2 t_{aa} t_{12}}{U} 
+\dfrac{t_{12} (t_{11}+t_{22}) }{U^{\prime}} 
\ 
\text{for} \ 
a^{\prime}=
a^{\prime\prime}= 
b^{\prime}=a, 
b^{\prime\prime}=\bar{a} \\
\dfrac{2 t_{aa} t_{12}}{U} 
+\dfrac{t_{12} (t_{11}+t_{22}) }{U^{\prime}} 
\ 
\text{for} \ 
a^{\prime}=
a^{\prime\prime}= 
b^{\prime\prime}=a, 
b^{\prime}=\bar{a} \\
\dfrac{2 t_{aa} t_{12}}{U} 
+ \dfrac{t_{12} (t_{11}+t_{22}) }{U^{\prime}} 
\ 
\text{for} \ 
a^{\prime}=
b^{\prime\prime}= 
b^{\prime}=a, 
a^{\prime\prime}=\bar{a} \\
\end{cases},\label{eq:J-form}
\end{align}
respectively. 
Note that 
the hopping integrals 
$t_{11}$, $t_{22}$, and $t_{12}(=t_{21})$ 
depend on the direction of $(\boldi-\boldj)$. 
The form of 
$J_{a^{\prime} a^{\prime\prime}, 
b^{\prime} b^{\prime\prime}}$ 
indicates that 
the number of SC attractive channels 
for electrons is enhanced by 
the coupling between spin and orbital 
degrees of freedom. 

At $T=T_{\textrm{c}}$, 
the SC gap function satisfies 
the following self-consistent equation: 
\begin{align}
&\Delta_{ab}(\boldk)
= 
\frac{1}{N} 
\textstyle\sum\limits_{\boldk^{\prime}} 
\textstyle\sum\limits_{a^{\prime}, b^{\prime}, 
a^{\prime\prime}, b^{\prime\prime}=1}^{2} 
\textstyle\sum\limits_{\alpha^{\prime}, \beta^{\prime}} 
V_{a a^{\prime}, b b^{\prime}}(\boldk - \boldk^{\prime})\notag\\
( &
\tanh{ \dfrac{ \tilde{\epsilon}_{\alpha^{\prime}}
(\boldk^{\prime}) }
{ 2 T_{\textrm{c}}}} 
+ 
\tanh{ \dfrac{ \tilde{\epsilon}_{\beta^{\prime}}
(-\boldk^{\prime}) }
{ 2 T_{\textrm{c}}}}
) 
\dfrac{ 
\Delta_{a^{\prime\prime} b^{\prime\prime}}(\boldk^{\prime}) 
}
{ 2 ( 
\tilde{\epsilon}_{\alpha^{\prime}}(\boldk^{\prime}) 
+ \tilde{\epsilon}_{\beta^{\prime}}(-\boldk^{\prime}) 
) 
}\notag\\
(&U_{\boldk^{\prime}})_{a^{\prime} \alpha^{\prime}} 
(U_{\boldk^{\prime}}^{\dagger})
_{\alpha^{\prime} a^{\prime\prime}} 
(U_{-\boldk^{\prime}})_{b^{\prime} \beta^{\prime}} 
(U_{-\boldk^{\prime}}^{\dagger})
_{\beta^{\prime} b^{\prime\prime}} . 
\label{eq:defDel2}
\end{align}
Note that 
the sum of the orbital indices 
$(a^{\prime},b^{\prime},
a^{\prime\prime},b^{\prime\prime})$ 
is restricted to 
the $d_{xz}$ and $d_{yz}$ orbitals, 
while there is no restriction for the sum of 
the band indices $(\alpha^{\prime},\beta^{\prime})$. 
The effect of the other three Fe $3d$-orbitals (i.e., 
$d_{xy}$, $d_{x^{2}-y^{2}}$, and $d_{z^{2}}$) 
is included in the energy dispersions and unitary matrices, 
since 
the $d_{xz}$ and $d_{yz}$ orbitals have finite hybridizations 
with the other three Fe $3d$-orbitals. 
$T_{\textrm{c}}$ can be estimated 
by the following linearized gap equation: 
\begin{align}
&\lambda_{\textrm{e}} \Delta_{ab}(\boldk)
= 
\frac{1}{N} 
\textstyle\sum\limits_{\boldk^{\prime}} 
\textstyle\sum\limits_{a^{\prime}, b^{\prime}, 
a^{\prime\prime}, b^{\prime\prime}=1}^{2} 
\textstyle\sum\limits_{\alpha^{\prime}, \beta^{\prime}} 
V_{a a^{\prime}, b b^{\prime}}(\boldk - \boldk^{\prime})\notag\\
( &
\tanh{ \dfrac{ \tilde{\epsilon}_{\alpha^{\prime}}
(\boldk^{\prime}) }
{ 2 T }} 
+ 
\tanh{ \dfrac{ \tilde{\epsilon}_{\beta^{\prime}}
(-\boldk^{\prime}) }
{ 2 T }}
) 
\dfrac{ 
\Delta_{a^{\prime\prime} b^{\prime\prime}}(\boldk^{\prime}) 
}
{ 2 ( 
\tilde{\epsilon}_{\alpha^{\prime}}(\boldk^{\prime}) 
+ \tilde{\epsilon}_{\beta^{\prime}}(-\boldk^{\prime}) 
) 
}\notag\\
(&U_{\boldk^{\prime}})_{a^{\prime} \alpha^{\prime}} 
(U_{\boldk^{\prime}}^{\dagger})
_{\alpha^{\prime} a^{\prime\prime}} 
(U_{-\boldk^{\prime}})_{b^{\prime} \beta^{\prime}} 
(U_{-\boldk^{\prime}}^{\dagger})
_{\beta^{\prime} b^{\prime\prime}} , 
\label{eq:defDel3}
\end{align} 
where 
$T_{\textrm{c}}$ corresponds 
to the temperature 
at which the eigenvalue $\lambda_{\textrm{e}}$ 
becomes unity.  

We remark on the symmetry of the pair amplitude 
in terms of the orbital degree of freedom. 
The pair amplitude satisfies 
\begin{align}
\langle 
c_{-\boldk a\downarrow} 
c_{\boldk b \uparrow} 
\rangle
=
\pm 
\langle 
c_{-\boldk b \downarrow} 
c_{\boldk a \uparrow} 
\rangle  \ \ \ \ \textrm{for} \ \ a\neq b \ ,
\end{align}
where the upper and lower signs correspond to 
orbital-ferromagnetic and orbital-AF pairings, 
respectively. 
According to the form of the interaction 
given by eq. (\ref{eq:J-form}), 
the attractive interaction is dominant 
in the case of 
orbital-ferromagnetic pairing. 
In the following calculations, 
we thus consider only 
the orbital-ferromagnetic spin-singlet pairing. 

\section{Results}
In this section, 
we show the results of numerical calculations 
using the MFA. 
We take 254$\times$254 meshes 
in the unfolded Brillouin zone (BZ) 
and set 
$n_{\textrm{e}}=$6.24, 
$\varepsilon_{\textrm{c}}=$0.01 eV, 
$(U, U^{\prime}, J_{\textrm{H}})
=(1, 0.8, 0.1)$ (eV), 
and the temperature $T=$0.003 eV.  
For convenience, 
the SC gap function 
is always normalized. 
According to the 2-D tight-binding calculation 
in the tetragonal phase,~\cite{Kuroki} 
hopping integrals for the $d_{xz}$ and $d_{yz}$ orbitals 
are given by 
$(t_{11}^{\textrm{n.n.}})_{\parallel x}
=(t_{22}^{\textrm{n.n.}})_{\parallel y}
=$ 0.08 eV, 
$(t_{11}^{\textrm{n.n.}})_{\parallel y}
=
(t_{22}^{\textrm{n.n.}})_{\parallel x}
=$ 0.34 eV,
$t_{12}^{\textrm{n.n.n.}}=0.00$eV,  
$t_{11}^{\textrm{n.n.n.}}
=
t_{22}^{\textrm{n.n.n.}}
=$ -0.24 eV, 
and 
$t_{12}^{\textrm{n.n.n.}}
=$ -0.09 eV. 
(e.g., $(t_{11}^{\textrm{n.n.}})_{\parallel x}$ 
denotes $(t_{11}^{\boldi,\boldj})$ for 
$(\boldi-\boldj)=(\pm 1, 0)$, 
etc.) 
Hereafter, 
the unit of energy is taken as $1$ eV. 

We remark on the validity of 
our choice of $J_{\textrm{H}}$ and $U$. 
($U^{\prime}$ is given by 
the relation $U^{\prime}=U-2J_{\textrm{H}}$.) 
A recent study on 
the five-orbital Hubbard model within the MFA 
shows that 
the small magnetic moment in the parent compounds 
can be understood 
only in the case of small values of $J_{\textrm{H}}/U$.~\cite{HubbardAF} 
This is also supported by Gutzwiller approximation studies on 
the two- and three-orbital Hubbard models.~\cite{KuboAF2,3-AFM} 
In accordance with these results, 
we use small values of $J_{\textrm{H}}/U$ 
to discuss the SC state 
in the vicinity of the AF state. 

\subsection{Renormalized band structure}
\begin{table}[tb]
\vspace{-5pt}
\caption{Occupation numbers 
of electrons of the five Fe $3d$ orbitals 
in the original 
and renormalized band structures 
in the tetragonal phase.}
\label{tab:ne}
\begin{center}
\begin{tabular}{ccc}
\\
\hline
\multicolumn{1}{c}{ \ \ Orbital \ } & 
\multicolumn{1}{c}{ \ Original \ } & 
\multicolumn{1}{c}{ \ Renormalized \ \ \ } \\
\hline
$d_{xz}$ 
& 1.19 & 1.31 \\
$d_{yz}$ 
& 1.19 & 1.31 \\
$d_{xy}$ 
& 1.14 & 1.20 \\
$d_{x^{2}-y^{2}}$ 
& 0.98 & 0.98 \\
$d_{z^{2}}$ 
& 1.49 & 1.46 \\
\hline
\end{tabular}
\end{center}
\end{table}
%
\begin{figure}[tb]
(a)
\hspace{100pt}
(b)

\begin{tabular}{cc}
{
\includegraphics[width=24mm, bb= 10 0 90 100]{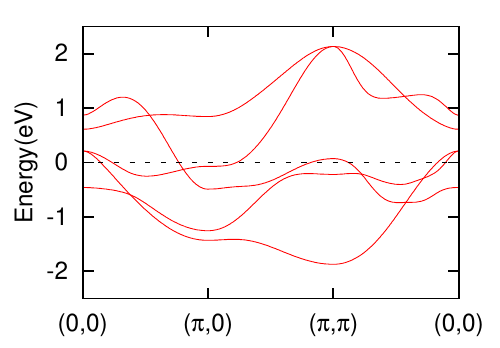}}& 
{\hspace{30pt}
\includegraphics[width=24mm, bb= 0 0 80 100]{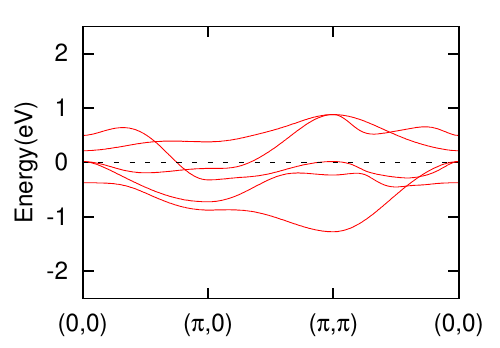}}\\[-7pt]
\end{tabular}
\caption{(Color online) 
(a) Original band structure and 
(b) renormalized band structure 
in the tetragonal phase. 
The dashed lines represent 
the chemical potentials. }
\label{fig:band-tetra}
\end{figure}
\begin{figure}[tb]
\vspace{-10pt}
\hspace{24pt}
(a)
\hspace{86pt}
(b)

\vspace{-22pt}
\begin{tabular}{cc}
{
\includegraphics[width=32mm, bb= 20 0 100 100]
{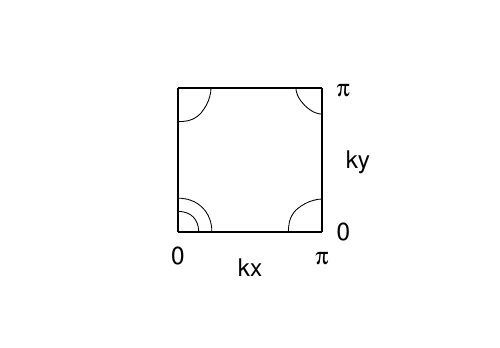}}&
{
\includegraphics[width=32mm, bb= 18 0 98 100]{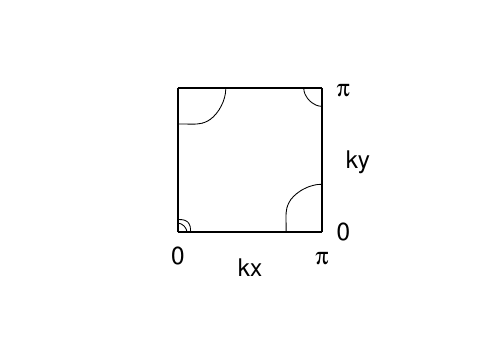}}\\[-25pt]
\end{tabular}
\caption{
FSs 
for the (a) original and 
(b) renormalized band structures 
in the tetragonal phase. }
\label{fig:FS-tetra}
\vspace{-5pt}
\end{figure}
\begin{figure}[tb]
(a)
\hspace{100pt}
(b)

\vspace{6pt}
\begin{tabular}{cc}
{
\includegraphics[width=22mm, bb= 10 0 90 100]
{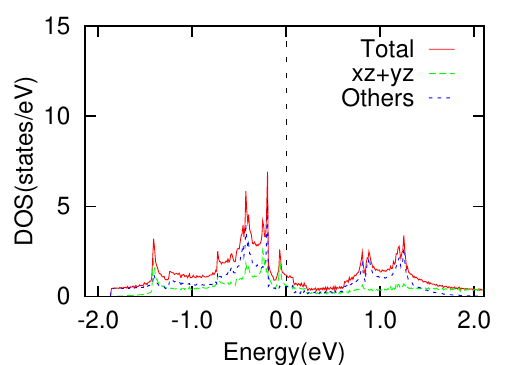}}&
{\hspace{36pt}
\includegraphics[width=22mm, bb= 0 0 80 100]{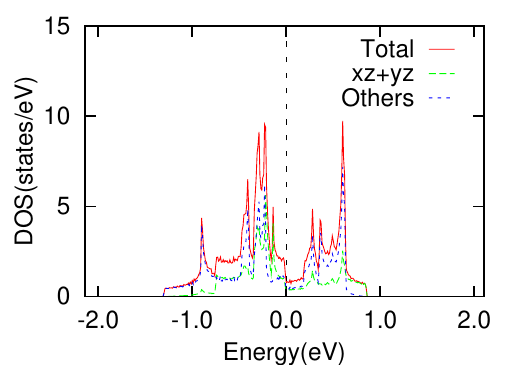}}\\[-5pt]
\end{tabular}
\caption{(Color online) 
DOSs for the (a) original 
and (b) renormalized band structures 
in the tetragonal phase. 
The dashed lines represent 
the chemical potentials. }
\label{fig:DOS-tetra}
\vspace{-5pt}
\end{figure}
\begin{figure}[tb]
\hspace{14pt}
(a)
\hspace{70pt}
(b)

\vspace{-7pt}
\begin{tabular}{cc}
{
\includegraphics[width=32mm, bb= 30 10 110 90]
{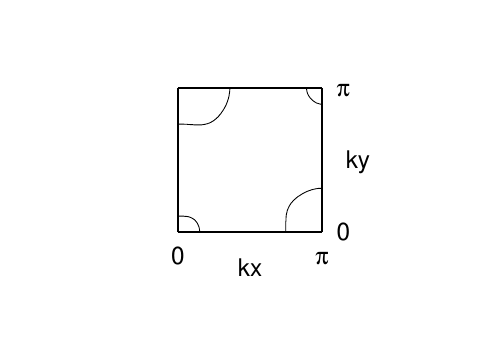}}&
{
\includegraphics[width=22mm, bb= 0 0 80 100]{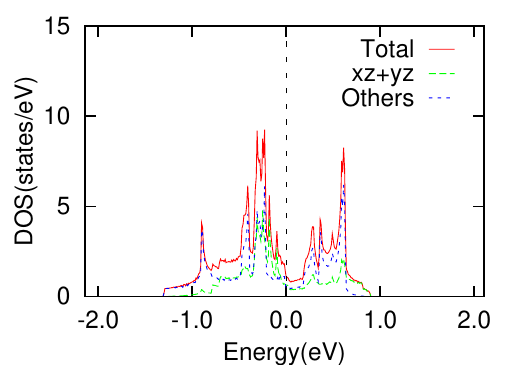}}\\[-5pt]
\end{tabular}
\caption{(Color online) 
(a) FS and (b) DOS for $\Delta_{\textrm{ortho.}}=0.08$. 
The dashed line represents 
the chemical potential. }
\label{fig:FS-DOS-ortho}
\vspace{-5pt}
\end{figure}
As described in \S 2.1, 
a renormalized band structure is constructed 
in order to take into account 
the relatively strong electron correlation 
for the $d_{xz}$ and $d_{yz}$ orbitals. 
By this procedure, 
$(n_{\textrm{e}})_{xz+yz}$ 
becomes larger than 
that in the original band structure 
and is equal to 2.62, 
which is close to 3 (Table \ref{tab:ne}). 

The obtained band structure, 
the corresponding FSs, and 
the density of states (DOS)  
in the tetragonal phase  
are shown in Figs. \ref{fig:band-tetra}-\ref{fig:DOS-tetra}, 
respectively. 
For comparison, 
the original band structure, 
the corresponding FSs, 
and the DOS are
also depicted in the figures. 

Similarly, 
we calculate the FSs and DOS 
in the orthorhombic phase. 
As a typical case,
the results for $\Delta_{\textrm{ortho.}}=0.08$ 
are shown in Fig. \ref{fig:FS-DOS-ortho}. 
The main effects of 
orthorhombic distortion 
are i) the disappearance of one of the FSs around 
the $\Gamma$-point [i.e., $\boldk=(0,0)$], 
and ii) the reduction of the DOS  
in the vicinity of the Fermi level. 

\subsection{Eigenvalue of the linearized gap equation}
\begin{figure}[tb]
\vspace{-30pt}
\begin{center}
\includegraphics[width=50mm, bb= 20 0 110 100]
{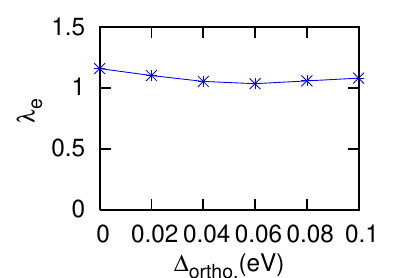}
\end{center}
\vspace{-4pt}
\caption{(Color online) 
Eigenvalue of the linearized gap 
equation, $\lambda_{\textrm{e}}$, 
for $0\leq \Delta_{\textrm{ortho.}} \leq 0.1$.  }
\label{fig:lambda}
\end{figure}
%
\begin{figure}[tb]
\vspace{-6pt}
(a)
\hspace{100pt}
(b)

\vspace{-24pt}
\subfigure{
\includegraphics[width=30mm, bb= 2 0 82 100]
{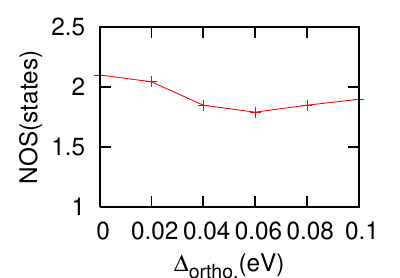}}
\subfigure{\hspace{26pt}
\includegraphics[width=30mm, bb= 0 0 80 100]{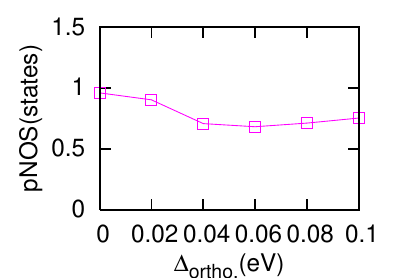}}
\vspace{-10pt}
\caption{(Color online) 
(a) NOS 
and (b) pNOS 
for $d_{xz}$ and $d_{yz}$ orbitals 
in the vicinity of the Fermi level 
as a function of $\Delta_{\textrm{ortho.}}$. }
\label{fig:frac}
\vspace{-5pt}
\end{figure}
\begin{figure}[tb]
\hspace{24pt}
(a)
\hspace{86pt}
(b)

\vspace{-22pt}
\begin{tabular}{cc}
{
\includegraphics[width=32mm, bb= 20 0 100 100]
{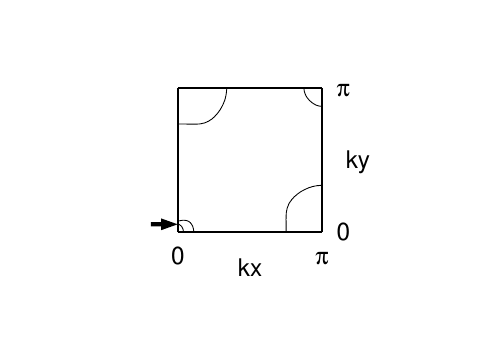}}&
{
\includegraphics[width=32mm, bb= 18 0 98 100]{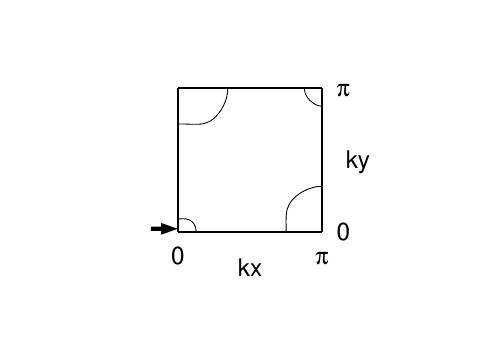}}\\[-25pt]
\end{tabular}
\caption{
FSs for $\Delta_{\textrm{ortho.}}=$
(a) 0.02, and 
(b) 0.04. 
The main changes in the FSs are shown by arrows. 
}
\label{fig:FS-dev}
\vspace{-5pt}
\end{figure}

\begin{figure}[tb]
(a)
\hspace{110pt}
(b)

\vspace{-54pt}
\subfigure{\hspace{-20pt}
\includegraphics[width=34mm, bb= 40 0 120 100]
{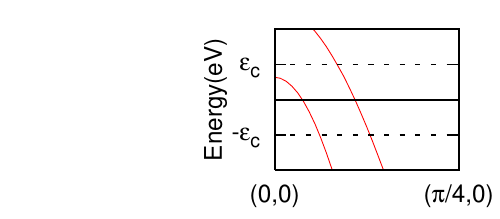}}
\subfigure{\hspace{-10pt}
\includegraphics[width=34mm, bb= 10 0 90 100]{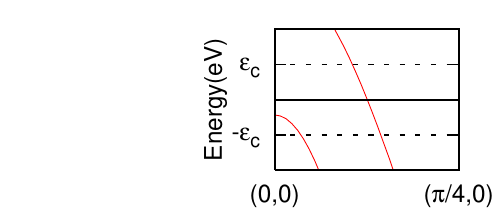}}
\vspace{-10pt}
\caption{(Color online)
Band structures 
around the $\Gamma$ point 
for $\Delta_{\textrm{ortho.}}=$
(a) 0.02, and 
(b) 0.04. 
Solid lines and dashed lines correspond to 
the chemical potentials and 
cutoff energies $\varepsilon_{\textrm{c}}$($=0.01$), 
respectively. }
\label{fig:band-dev}
\vspace{-10pt}
\end{figure}
We investigate the eigenvalue, 
$\lambda_{\textrm{e}}$, 
of the linearized gap equation, eq. (\ref{eq:defDel3}), 
for $0\leq \Delta_{\textrm{ortho.}} \leq 0.1$. 
Figure \ref{fig:lambda} shows the obtained eigenvalue, 
$\lambda_{\textrm{e}}$, as a function of 
$\Delta_{\textrm{ortho.}}$. 
We find that 
$\lambda_{\textrm{e}}$ decreases 
as $\Delta_{\textrm{ortho.}}$ increases. 
This result means that 
$T_{\textrm{c}}$ is highest in the tetragonal phase. 
In order to investigate the physical origin 
of this behavior of $\lambda_{\textrm{e}}$, 
we calculate 
the number of states (NOS) near the Fermi level, 
which is defined as 
the DOS integrated from $-\varepsilon_{\textrm{c}}$ 
to $\varepsilon_{\textrm{c}}$. 
Figure \ref{fig:frac} shows the total NOS and 
the partial number of states (pNOS) 
for the $d_{xz}$ and $d_{yz}$ orbitals. 
Comparing Fig. \ref{fig:lambda} with Fig. \ref{fig:frac}, 
we consider that 
the physical origin of the behavior of 
$\lambda_{\textrm{e}}$ 
is the variation of the pDOS 
for the $d_{xz}$ and $d_{yz}$ orbitals. 
This result indicates that 
energy-splitting 
due to the orthorhombic distortion 
plays an important role 
in controlling $T_{\textrm{c}}$. 
A similar conclusion was also obtained 
theoretically for CeCoIn$_{5}$.~\cite{Gap-symmetry} 

Figure \ref{fig:lambda} also shows that 
$\lambda_{\textrm{e}}$ decreases 
with increasing $\Delta_{\textrm{ortho.}}$ 
for small values of $\Delta_{\textrm{ortho.}}$, 
while it does not change greatly 
for $\Delta_{\textrm{ortho.}}>0.05$. 
We speculate that 
this difference originates from the existence  
of a small hole pocket around the $\Gamma$-point. 
Figures \ref{fig:FS-dev} and \ref{fig:band-dev} 
show the main changes in the FSs  
for $\Delta_{\textrm{ortho.}}=0.02$ and $0.04$ 
and 
the corresponding band structures 
near the $\Gamma$-point, 
respectively. 
The arrows in Fig. \ref{fig:FS-dev} represent the changes in the FSs. 
One of the small hole pockets 
around the $\Gamma$-point 
seems to disappear near $\Delta_{\textrm{ortho.}}=0.04$. 
We consider that 
the small hole pocket enhances $\lambda_{\textrm{e}}$. 
When it vanishes above $\Delta_{\textrm{ortho.}}=0.04$, 
$\lambda_{\textrm{e}}$ does not change further. 
A similar effect has been observed in a previous study 
for a system with multiple FSs.~\cite{KurokiPRB} 

\subsection{SC gap function and pairing symmetry}
\begin{figure}[tb]
\vspace{-30pt}
\begin{center}
\includegraphics[width=50mm, bb= 20 0 110 100]
{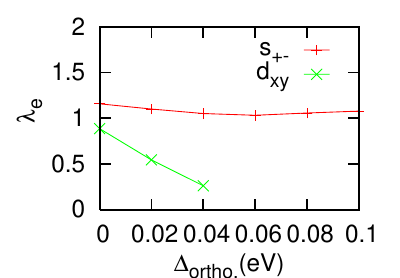}
\end{center}
\vspace{-4pt}
\caption{(Color online) 
Eigenvalue of the linearized 
gap equation $\lambda_{\textrm{e}}$ 
with $s$-wave and $d$-wave symmetries 
for $0\leq \Delta_{\textrm{ortho.}} \leq 0.1$. }
\label{fig:various-lambda}
\end{figure}
\begin{figure}[tb]
(a) $\Delta_{11}$
\hspace{84pt}
(b) $\Delta_{22}$

\vspace{-10pt}
\begin{tabular}{cc}
{
\includegraphics[width=32mm, bb= 30 0 110 100]{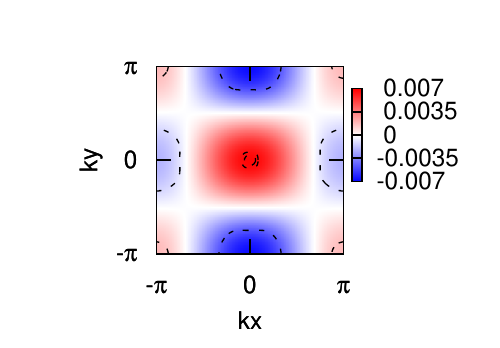}}& 
{\hspace{20pt}
\includegraphics[width=32mm, bb= 30 0 110 100]{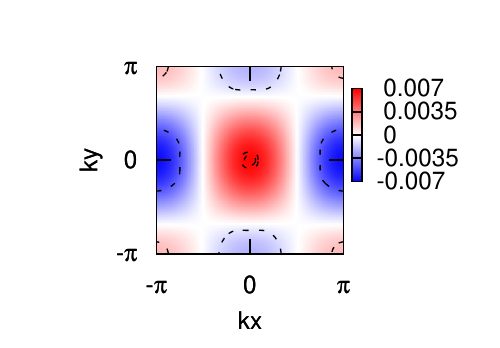}}\\[-7pt]
\end{tabular}
\caption{(Color online) 
SC gap functions $\Delta_{11}$ and $\Delta_{22}$ 
for $\Delta_{\textrm{ortho.}}=0$. 
The dashed lines represent the FSs for $\Delta_{\textrm{ortho.}}=0$. 
}
\label{fig:gap-tetra}
\end{figure}
\begin{figure}[tb]
(a) $\Delta_{11}$
\hspace{84pt}
(b) $\Delta_{22}$

\vspace{-10pt}
\begin{tabular}{cc}
{
\includegraphics[width=32mm, bb= 30 0 110 100]{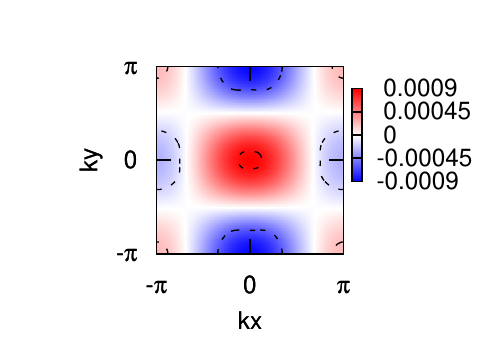}}& 
{\hspace{20pt}
\includegraphics[width=32mm, bb= 30 0 110 100]{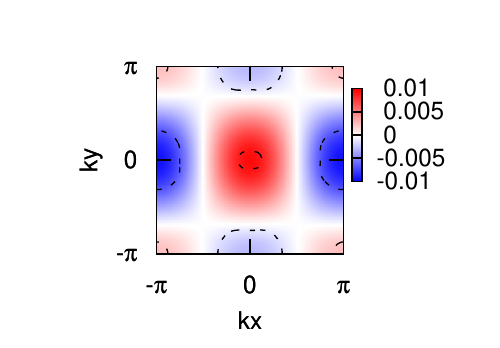}}\\[-7pt]
\end{tabular}
\caption{(Color online) 
SC gap functions $\Delta_{11}$ and $\Delta_{22}$ 
for $\Delta_{\textrm{ortho.}}=0.04$. 
The dashed lines represent the FSs for $\Delta_{\textrm{ortho.}}=0.04$. 
}
\label{fig:gap-ortho}
\end{figure}
\begin{figure}[tb]
\hspace{118pt}
$\Delta_{11/22}$
\vspace{-20pt}
\begin{center}
\includegraphics[width=40mm, bb= 24 0 114 100]
{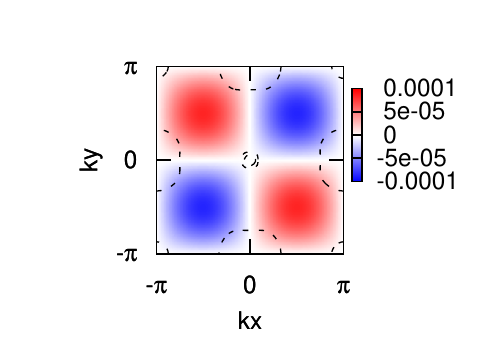}
\end{center}
\vspace{-10pt}
\caption{(Color online) 
SC gap functions $\Delta_{11/22}$ 
for a $d_{xy}$-wave pairing 
at $\Delta_{\textrm{ortho.}}=0$. 
The dashed lines represent the FSs for $\Delta_{\textrm{ortho.}}=0$. 
}
\label{fig:gap-dxy}
\end{figure}
\begin{figure}[tb]
(a)
\hspace{102pt}
(b)

\vspace{-24pt}
\subfigure{
\includegraphics[width=30mm, bb= 14 0 94 100]
{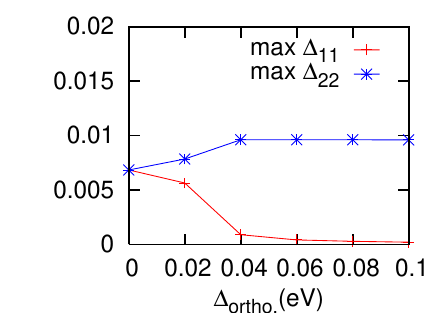}}
\subfigure{\hspace{38pt}
\includegraphics[width=33mm, bb= 10 0 90 100]{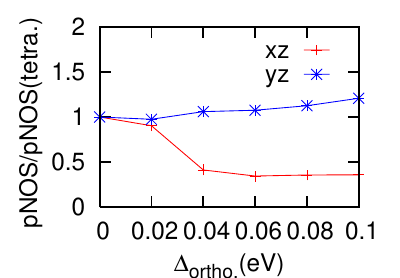}}
\vspace{-10pt}
\caption{(Color online) 
(a) Maximum values of orbital-diagonal SC gap functions 
for $s_{+-}$-wave symmetry, 
max $\Delta_{11}$ and max $\Delta_{22}$, 
as a function 
of $\Delta_{\textrm{ortho.}}$ 
and (b) pNOSs for the $d_{xz}$ and $d_{yz}$ orbitals 
normalized by those in the tetragonal phase. 
}
\label{fig:gap-dev}
\vspace{-5pt}
\end{figure}
We now turn to 
the SC gap function 
and its pairing symmetry. 
Before we show the results of numerical calculations, 
we remark on the method used to determine 
the pairing symmetry for the multiorbital SC state 
within the linearized gap equation.~\cite{Gap-symmetry} 
For the multiorbital SC state, 
there are two components of the SC gap function: 
orbital-diagonal and orbital-off-diagonal components. 
The pairing symmetry is determined 
from the orbital-diagonal SC gap function 
since it belongs to 
the same irreducible representation of the SC state. 
Hereafter, 
we show only the orbital-diagonal SC gap functions. 

First, 
we investigate the most dominant pairing symmetry. 
As shown in Fig. \ref{fig:various-lambda}, 
an unconventional fully gapped $s_{+-}$-wave pairing 
is the most stable 
for $0\leq \Delta_{\textrm{ortho.}}\leq 0.1$. 
As typical forms of the SC gap functions, 
$\Delta_{11}$ and $\Delta_{22}$ in the tetragonal and orthorhombic phases 
are shown in Figs. \ref{fig:gap-tetra} and 
\ref{fig:gap-ortho}, respectively.  
At $\Delta_{\textrm{ortho.}}=0$, 
the second most stable pairing symmetry is 
a $d_{xy}$-wave pairing 
whose form is depicted in Fig. \ref{fig:gap-dxy}. 
Note that 
the values of $\lambda_{\textrm{e}}$ for 
other $d$-wave pairings (e.g., 
the $d_{x^{2}-y^{2}}$-wave pairing) 
are smaller than that for the $d_{xy}$-wave pairing. 

Let us discuss the condition required 
to stabilize the $d_{xy}$-wave pairing. 
At $\Delta_{\textrm {ortho.}}=0$, 
the FS around the $\textrm{X}$- or $\textrm{Y}$-point [i.e., 
$(\pi,0)$ or $(0,\pi)$] 
does not have components 
of the $d_{xz}$ and $d_{yz}$ orbitals 
along the $k_{x}$- or $k_{y}$-direction, 
respectively.~\cite{KurokiPRB} 
Therefore, 
the energy cost is small even if a line node appears 
along the $k_{x}$- and $k_{y}$-axes. 
This is why the $d_{xy}$-wave 
pairing is stabilized. 
However, 
the rapid drop in $\lambda_{\textrm{e}}$ 
for the $d_{xy}$-wave pairing 
shown in Fig. \ref{fig:various-lambda} 
may be due to a subtle balance of the FSs 
around the $\Gamma$-point. 
 
Finally, 
we investigate 
the anisotropy of the SC gap function 
in the orthorhombic phase. 
Figure \ref{fig:gap-dev}(a) shows the maximum values 
of the SC gap functions, 
$\Delta_{11}$ and $\Delta_{22}$, 
as a function of $\Delta_{\textrm{ortho.}}$. 
When $\Delta_{\textrm{ortho.}}$ 
is on the order of 10$\%$ of the renormalized 
bandwidth ($\Delta_{\textrm{ortho.}}=$0.1), 
the ratio of max$\Delta_{11}$ to max$\Delta_{22}$ 
becomes 1$:$1000. 
Large anisotropy is thus expected for the SC state 
in the orthorhombic phase. 
With increasing $\Delta_{\textrm{ortho.}}$, 
the amplitude of $\Delta_{11}$ decreases rapidly, 
while that of $\Delta_{22}$ increases. [The SC gap 
functions for the $s_{+-}$-wave pairing 
become maximum at $\boldk=(0,0)$.] 
This behavior 
can be understood from the pNOS for the $d_{xz}$ and $d_{yz}$ orbitals 
shown in Fig. \ref{fig:gap-dev}(b). 
We can see that 
$\Delta_{\textrm{ortho.}}$ markedly affects 
the pNOS for the $d_{xz}$ orbital, 
while it does not greatly affect that for the $d_{yz}$ orbital. 
This is because the energy level 
of the $d_{xz}$ orbital decreases to below the Fermi level. 
This variation of pNOSs 
leads to the decrease in $\Delta_{11}$ 
and increase of $\Delta_{22}$ 
as $\Delta_{\textrm{ortho.}}$ increases. 

Since the experimentally observed SC gap function 
is the sum of the orbital-diagonal and orbital-off-diagonal 
components, 
the SC gap function becomes anisotropic in the case of  
different values of $\Delta_{11}$ and $\Delta_{22}$. 
Figure \ref{fig:gap-dev} indicates that 
the SC gap function around the X-point becomes larger 
than that around the Y-point for the SC state in the orthorhombic phase, 
since $\Delta_{22}$ is dominant in the orthorhombic phase 
and has this anisotropy (Fig. \ref{fig:gap-ortho}). 
We thus expect a large anisotropy of the SC gap function 
in the orthorhombic phase, which can be detected 
by ARPES 
measurement of the SC state in the orthorhombic phase. 
As we shall discuss in \S 4, 
this anisotropy is also expected 
in a SC state coexisting with a nematic state. 

\section{Discussion}
First, 
we discuss the validity of neglecting  
the anisotropy of hopping integrals 
due to the orthorhombic distortion. 
According to a recent tight-binding calculation 
for the parent compound 
of LaFeAsO,~\cite{Pickett} 
there is little significant difference 
in the hopping integrals 
for the $d_{xz}$ and $d_{yz}$ orbitals. 
Since we focus only on the region 
where the orthorhombic distortion 
is small ($\Delta_{\textrm{ortho.}}\leq 0.1$), 
the effect of the anisotropy on the hopping integrals 
is negligible. 
In other words, 
the effect of orthorhombic distortion 
on the energy-splitting 
will be more important than 
that on the hopping integrals 
in the presence of a small orthorhombic distortion. 
It is reasonable to consider only 
a small orthorhombic distortion 
since we consider 
the superconductivity 
in the doped case where 
the effect of orthorhombic distortion 
is suppressed. 

We address the role of the spin-orbital coupling 
in determining the SC state. 
As shown in \S 2, 
the degeneracy of orbitals gives 
additional degrees of freedom which enhance 
the number of attractive channels 
for Cooper pairing. 
On the other hand, 
the orbital degree of freedom also 
has a depairing effect 
on superconductivity. 
In the present MFA, 
we have not clarified the depairing effect. 

In iron-based compounds, 
As ions are located asymmetrically 
with respect to the Fe plane.~\cite{Fe-review} 
Therefore, 
there are many hopping integrals between Fe $3d$ orbitals, 
which are not allowed without As ions. 
This increases the number of interactions 
in eq. (\ref{eq:KK}), 
which eventually increases the number of SC attractive channels. 
Note that 
these terms also enhance the magnetic instability; 
thus, they do not 
always favor the SC state. 

Next, 
we compare our theory with other previous theoretical studies. 
As shown in \S 1, 
there are two different candidates for the pairing symmetry of 
the iron-based superconductors, i.e., 
$s_{+-}$-pairing~\cite{Mazin, Kuroki,Nomura,Ikeda,fRG,Fuseya,
122-RPA,122-Kuroki} 
and $s_{++}$-pairing~\cite{Kontani1, Yanagi}. 
Our theory suggests the former pairing symmetry. 
Although the mechanism is not clear yet, 
we consider that 
the antiferromagnetism is closely related to the emergence 
of the superconductivity. 
Note that 
the quantum critical point 
of the antiferromagnetism has been observed 
for optimally doped 
$\textrm{Ba}\textrm{Fe}_{2}
(\textrm{As}_{1-x}\textrm{P}_{x})_{2}$ 
by a nuclear magnetic resonance 
(NMR) measurement.~\cite{P122NMRQCP} 
This result indicates that 
the antiferromagnetism plays an important role 
in the emergence of superconductivity for iron-based compounds. 

We remark on the main difference between our theory and 
spin-fluctuation-mediated pairing, 
whether the glue of the superconductivity 
exists or not. 
In this paper, 
the SC attractive force originates directly from 
the KK-type superexchange interactions, 
while in spin-fluctuation theory, 
it originates from the fluctuation using 
a wide range of frequency space. 
Our obtained results have only n.n. and n.n.n. SC gap functions. 
This is qualitatively consistent with the SC gap functions 
obtained in the spin-fluctuation-mediated pairing, 
since the latter can be reproduced 
by the n.n. and n.n.n. parameters in real space.~\cite{real-gap} 
The SC gap function observed in several ARPES measurements 
can also be fitted with the n.n. and n.n.n. parameters 
in real space.~\cite{ARPES122,ARPES122-2} 
Therefore, 
the behaviors of $\lambda_{\textrm{e}}$ 
and the anisotropy between $\Delta_{11}$ and $\Delta_{22}$ 
as a function of $\Delta_{\textrm{ortho.}}$ 
will be qualitatively general in iron-based superconductors. 

Next, 
we discuss the experimental correspondence 
for our assumption 
that the attractive interaction 
appears only among the special orbitals 
such as $d_{xz}$ and $d_{yz}$. 
This assumption is based on the idea 
that the AF state in parent compounds 
can be described 
by two localized orbitals.~\cite{FO-aniso1} 
Then, electrons in these orbitals 
induce superconductivity 
due to chemical doping and/or physical doping. 
We consider that 
this orbital-selective AF state does not contradict 
the experimental results for the metallic behavior, 
since the other orbitals have itinerant character. 
(A similar situation is expected 
in $\textrm{Ca}_{2-x}\textrm{Sr}_{x}
\textrm{Ru}\textrm{O}_{4}$.~\cite{OSS-Mott}) 
As shown in \S 1, 
there are several experimental results 
that support our assumption: 
the large incoherence of the electronic Raman spectrum 
for these orbitals,~\cite{122Raman-loc} 
and the smaller mobility of holes observed in 
Hall resistivity measurement using a two-carrier model~\cite{122Hall-loc}. 
Note that the hole mobility will mainly originate 
from the $d_{xz}$ and $d_{yz}$ orbitals, 
since the hole pockets of the FS usually consists of these orbitals. 
Moreover, 
theoretical calculations for FeSe 
using the dynamical mean field 
theory~\cite{Imada-DMFT,Max-Planck-DMFT} 
show that 
the renormalization factors for the $d_{xz}$ and $d_{yz}$ orbitals 
are small ($=0.28$) compared with 
those of $d_{z^{2}}$ and $d_{x^{2}-y^{2}}$. 
(Note, however, 
that the renormalization factor 
for the $d_{xy}$ orbital 
is also small.)
This is consistent with our assumption.  

In 
$\textrm{Ba}_{1-x}\textrm{K}_{x}
\textrm{Fe}_{2}\textrm{As}_{2}$ 
and 
$\textrm{Ba}\textrm{Fe}_{2}
(\textrm{As}_{1-x}\textrm{P}_{x})_{2}$, 
however, 
the ARPES measurement~\cite{ARPES-z2} 
indicates that the $d_{z^{2}}$ orbital 
contributes to the FS, 
and  
the SC gap of the $d_{z^{2}}$ orbital is 
as large as that of the other orbitals. 
In this case, 
it is necessary to include the $d_{z^{2}}$ orbital, 
although 
the $d_{z^{2}}$ orbital can be a passive orbital 
for superconductivity.  
   
We next address the possibility of superconductivity 
in the orthorhombic phase. 
Recently, 
the emergence of superconductivity 
in the orthorhombic phase 
with nearly 100$\%$ volume fraction 
has been reported for 
single crystals of 
$\textrm{Ba}_{0.6}\textrm{K}_{0.4}
\textrm{Fe}_{2}\textrm{As}_{2}$,~\cite{ortho-SC2} 
although there have been some reports 
for the same compound 
that claim 
the inhomogeneous coexistence of 
SC and AF states.~\cite{K122NMR,K-122-inhomo} 
According to ref. 3, 
$T_{\textrm{c}}$ increases as 
the orthorhombicity 
$\delta=(a-b)/(a+b)$ decreases. 
This is consistent with 
our results. 
As discussed in \S 3.2, 
the decrease in $T_{\textrm{c}}$ 
due to the small orthorhombic distortion 
can be understood 
as the reduction of the pDOS 
for the $d_{xz}$ and $d_{yz}$ orbitals. 
Furthermore, 
in FeSe$_{0.94}$ under high pressure, 
the microscopic coexistence of 
SC and AF states in the region of higher pressure
was observed by the muon-spin rotation measurement.~\cite{11coex} 
The SC state may also exist in the orthorhombic phase of 
$\textrm{Ba}(\textrm{Fe}_{1-x}\textrm{Co}_{x})_{2}
\textrm{As}_{2}$.~\cite{122PD,Co122-ortho.} 

On the other hand, 
in SrFe$_{2}$As$_{2}$ under high pressure, 
the NMR spectrum for the SC-dominant phase is 
almost equal to that 
for the tetragonal paramagnetic state.~\cite{122Kitagawa} 
This indicates that 
superconductivity is not realized in the orthorhombic phase. 
Although more detailed analysis in the presence of the AF order is needed, 
we consider that 
our effective model can qualitatively 
describe the properties 
of the SC state 
if superconductivity is realized in the orthorhombic phase. 

Finally, 
we remark on the nematic state 
observed in the 122 system~\cite{aniso-resis,Shen}, 
since the SC state coexisting with the nematic state 
is nearly the same as that in the orthorhombic phase. 
The nematic state is characterized by 
the lowering of rotational symmetry 
without any changes in the crystal structure. 
There have been some proposals of 
the emergence of the nematic state 
in strongly correlated systems 
such as cuprate~\cite{Cuprate-nematic} 
and Sr$_{3}$Ru$_{2}$O$_{7}$~\cite{Ru327}.  
In the iron-based compounds, 
its emergence has been proposed 
on the basis of 
experimental results 
for $\textrm{Ba}(\textrm{Fe}_{1-x}\textrm{Co}_{x})_{2}
\textrm{As}_{2}$~\cite{aniso-resis,Shen} 
and proposed 
theoretically.~\cite{nematic1111,STM-theory,Eremin-nematic,
Mazin-nematic}  
Although the effect of nematicity on the SC state has not 
been clarified, 
we propose that 
the anisotropy of the SC gap function, 
which is similar to that shown in Fig. \ref{fig:gap-dev}(a), 
i.e., a difference 
between the SC gap functions around the X- and Y-points, 
can be observed 
if the SC state coexists with the nematic state. 

\section{Summary}
We investigated the eigenvalue, $\lambda_{\textrm{e}}$, 
of the linearized gap equation 
and the pairing symmetry 
in a model for iron-based superconductors 
in both tetragonal and 
orthorhombic phases 
on the basis of the MFA. 
Our effective model 
consists of the five-orbital kinetic energy, 
the orthorhombic CEF energy, 
and the two-orbital KK-type superexchange 
interaction~\cite{K-K}. 
The effect of orthorhombic distortion on 
electronic states 
was taken into consideration 
as the energy-splitting 
of $d_{xz}$ and $d_{yz}$ orbitals, 
$\Delta_{\textrm{ortho.}}$. 
In order to take account of the relatively strong 
electron correlation, 
the procedure of band renormalization was used. 
We found a decrease 
in $\lambda_{\textrm{e}}$ 
accompanied with a reduction of 
the pNOS of the $d_{xz}$ and $d_{yz}$ orbitals 
near the Fermi level as $\Delta_{\textrm{ortho.}}$ increases. 
This result is consistent with the results of 
a recent experiment.~\cite{ortho-SC2} 
The fully gapped $s_{+-}$-wave symmetry pairing 
is predominant 
in both tetragonal and orthorhombic phases. 
$\lambda_{\textrm{e}}$ for the $d_{xy}$-wave pairing, 
which is the second most dominant symmetry, 
rapidly decreases 
as $\Delta_{\textrm{ortho.}}$ increases. 
The highest $T_{\textrm{c}}$ 
for the fully gapped $s_{+-}$-wave pairing 
is obtained in the tetragonal phase. 
The SC gap function 
for the fully gapped $s_{+-}$-wave pairing 
becomes anisotropic in the orthorhombic phase 
due to the crystal symmetry. 
We found large anisotropy 
of the SC gap function 
in the orthorhombic phase, 
$\textrm{max}\Delta_{11}$:$\textrm{max}\Delta_{22}
\sim$1:1000, 
although $\Delta_{\textrm{ortho.}}$ 
is on the order on 10$\%$ of the renormalized bandwidth. 
Anisotropy originating from the 
difference in the values of the dominant orbital-diagonal 
SC gap function (i.e., $\Delta_{22}$) 
around the X- and Y-points can be detected 
by ARPES measurement 
if the superconductivity appears 
in the orthorhombic phase or 
in the coexisting phase with the nematic state. 

We propose that the CEF energy 
plays an important role in controlling $T_{\textrm{c}}$ 
and the SC gap function for multiorbital superconductors, 
and consider that 
the orbital-selective superconductivity 
discussed in this paper will be universal 
for the structure-sensitive superconductivity 
observed in multiorbital systems. 

\begin{acknowledgments}
The authors would like to thank Y. Yanase 
and T. Kariyado 
for fruitful discussions and useful comments.
\end{acknowledgments}

\end{document}